\documentclass{epl}

\title{Hysteretic clustering in granular gas}
\author{Ko van der Weele, Devaraj van der Meer, Michel Versluis, \\ and Detlef Lohse}
\shortauthor{Ko van der Weele \etal}
\institute{ Department of Applied Physics and  
J.M. Burgers Centre for Fluid Dynamics -\\
University of Twente,
P.O. Box 217, 7500 AE Enschede, The Netherlands
}

\begin{document}

\maketitle

\begin{abstract}
Granular material is vibro-fluidized in $N=2$ and $N=3$ connected compartments, respectively. For sufficiently strong shaking the granular gas is equi-partitioned, but if the shaking intensity is lowered, the gas clusters in one compartment. The phase transition towards the clustered state is of 2nd order for $N=2$ and of 1st order for $N=3$. In particular, the latter is \textit{hysteretic}. The experimental findings are accounted for within a dynamical model that exactly has the above properties.
\end{abstract}

One of the characteristic features of a granular gas is its tendency
to spontaneously separate in dense and dilute regions
\cite{goldhirsch93,mcnamara94,du95,jaeger96pt,jaeger96,kudrolli97,kadanoff99}.
This makes granular gases fundamentally different from ordinary
molecular gases. The dynamics of granular material is of importance for many industrial
applications where it is brought into motion in order to sort,
transport, or process it. Here clustering usually is an unwanted
effect and any further understanding may yield a substantial
economic benefit.   

The tendency to form clusters can be traced back to the fact that the collisions between the granules are inelastic. Some energy is dissipated in every collision, which means that a relatively dense region (where the particles collide more often than elsewhere) will dissipate more energy, and thus become even denser, resulting in a cluster of slow particles. Vice versa, because the particle number is conserved, relatively dilute regions will become more dilute. The few particles in these regions are very rapid ones. In terms of the granular temperature, which goes as the mean squared velocity of the particles, the clustering phenomenon can also be interpreted as a separation in cold and hot regions, as if Maxwell's demon were at work \cite{eggers99}.

A striking illustration of the clustering phenomenon is provided by the Maxwell-demon experiment \cite{schlichting96}, consisting of a box divided into two compartments by a wall of a certain height, with a few hundred small beads in each compartment. The beads are brought in a gaseous state by shaking the system vertically. If the shaking is vigorous enough, the inelasticity of the gas is overpowered by the energy input into the system, and the beads divide themselves uniformly over the two compartments as in any ordinary molecular gas. But if the driving is lowered below a certain level, the beads cluster in one of the two compartments. We end up with a "cold" compartment containing a lot of beads (moving rather sluggishly, hardly able to jump over the wall anymore) and a "hot" compartment containing only a few (much more lively) beads. In equilibrium, the average particle flux from left to right equals that from right to left.

Such an asymmetric equilibrium can only be explained if the flux of particles from one compartment to the other is not a monotonously increasing function of the number of particles. Rather it must be a function that attains a maximum (at a certain number of particles) and thereafter decreases again. In agreement with this, and based on the kinetic theory for dilute granular gases \cite{jenkins83,jenkins86,mcnamara98}, Eggers proposed the following analytic approximation for the flux \cite{eggers99} (rewritten here in a form suited for an arbitrary number of $N$ connected compartments),
\begin{eqnarray}
\label{eq-flux}
F(n_{k})=An_{k}^{2}e^{-N^2Bn_{k}^{2}} ,
\end{eqnarray}
which is indeed a one-humped function of $n_{k}$, with a maximum at
$n_{k}=1/(N\sqrt{B})$. Here $n_{k}$ is the fraction of the particles in
the k-th box, normalized to $\sum n_{k} = 1$.

The factors $A$ and $B$ depend on the particle properties (such as their radius $r$, and the restitution coefficient $e$ of the interparticle collisions) and on experimental parameters such as the height $h$ of the wall and the frequency $f$ and amplitude $a$ of the driving. The factor $A$ determines the absolute rate of the flux, and can simply be incorporated in the time scale. The phase transition towards the clustering state is only determined by the factor $B$, which for a 2D gas of spherical disks takes the form \cite{eggers99}
\begin{eqnarray}
\label{eq-B}
B= 4 \pi g r^{2} (1-e)^{2} \frac{h}{(a f)^2} \left(\frac{P}{l
    N^2}\right)^2 ,
\end{eqnarray}
where $P$ is the total number of particles, and $l$ is the width of each box. Again, we have chosen a notation that anticipates the generalization to a row of $N$ equal boxes. For a given granular material ($r$ and $e$ fixed), $B$ can be raised either by increasing the value of $h$ or by decreasing the value of $af$, i.e., by reducing the driving.

\begin{figure}
\begin{center}
\includegraphics[width=0.85\textwidth]{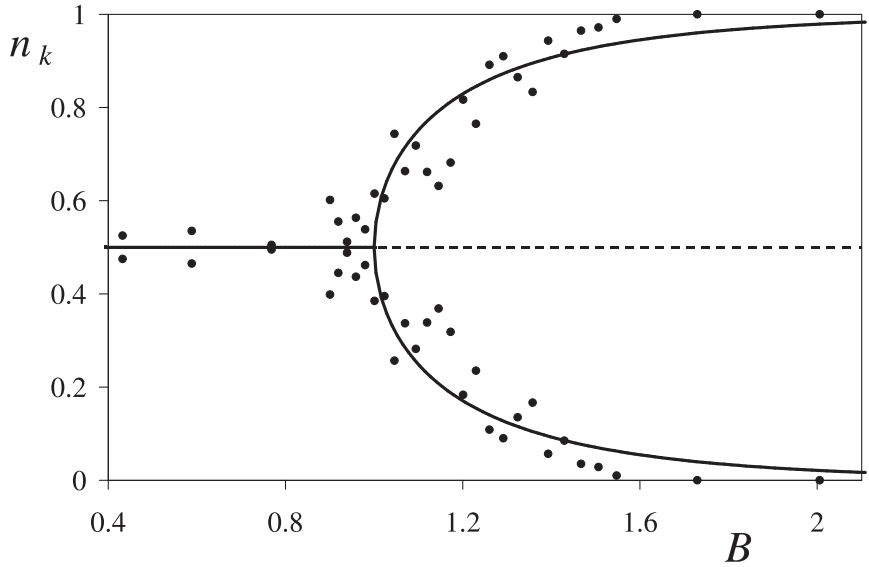}
\caption{Bifurcation diagram for the 2-box system ($k$ = 1, 2). 
The solid 
line represents stable and the  dashed line unstable equilibria of 
 the flux model. 
The dots are experimental measurements. In both cases, the transition to the clustering state is seen to take place via a pitchfork bifurcation.}
\label{fig-1}
\end{center}
\end{figure}

In order to check how well this flux equation works in practice we compared theory and experiment for the 2-box system. Theory says that the dynamics is given by the balance equation
\begin{eqnarray}
\label{eq-2box}
\frac{dn_1}{dt} = -F(n_1)+F(n_2) = -F(n_1)+F(1-n_1) ,
\end{eqnarray}
and in equilibrium one has $dn_1/dt=0$. One finds that for $B <
1$ there is only one (stable) equilibrium, namely the uniform $\{1/2,
1/2\}$ distribution. For $B > 1$ it becomes unstable and gives way to an asymmetric equilibrium. In other words, there is a symmetry breaking bifurcation at $B_{bif}=1$. In Fig.~\ref{fig-1} we have drawn the corresponding bifurcation diagram (cf. \cite{eggers99}).

For the experimental verification of this diagram we put 600 glass
beads ($r$ = 1.25 mm, $e$ = 0.97) in a cylindrical perspex tube of inner radius
27.5 mm, divided into two equal compartments by a wall of height $h$ =
23.0 mm. The tube was mounted on a shaker with an adjustable frequency
and amplitude, so that $B$ could be varied. In the present experiments
(for a 3D gas, with $B \propto 1/f^2$ still holding) this was done by varying
 of the frequency $f$, at a fixed value of $a$ = 6.5 mm. The measurements are included as solid dots in Fig.~\ref{fig-1}. Clearly, theory and experiment agree on the fact that the clustering transition takes place via a pitchfork bifurcation. It may be noted that the experimental curves bend towards the $\{1,0\}$ distribution earlier than predicted by the theory. This is due to the fact that the beads were counted not in the dynamical situation, but \textit{after} the shaker had been turned off and the particles had come to rest. During the final bounces, the fast particles still jump over the wall (from the empty compartment into the full one), whereas jumps in the opposite direction hardly occur at all. The result is that a dynamical equilibrium of say $\{0.97, 0.03\}$ is turned into a static state $\{1, 0\}$, and that is what we count.

\begin{figure}
\begin{center}
\includegraphics[width=0.85\textwidth]{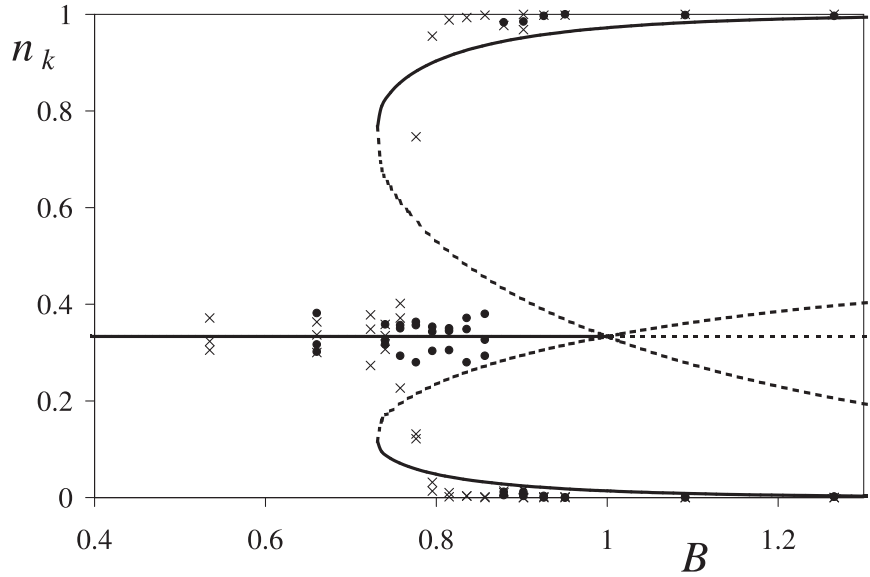}
\caption{Hysteretic
bifurcation diagram for the cyclic 3-box system ($k = 1, 2, 3$). 
The solid 
lines represent stable and the  dashed lines unstable equilibria of 
 the flux model. The bifurcations in this model take place at
$B_{bif}=1$ for increasing $B$ and at $B_{sn}=0.73$ for decreasing $B$. 
The dots and crosses are experimental measurements: The
dots for measurements that were started from the uniform distribution $\{1/3, 1/3, 1/3\}$ and the crosses for those that were started from a single peaked distribution. 
}
\label{fig-2}
\end{center}
\end{figure}

Now we generalize the experiment by taking not two, but three cyclic
compartments ($N = 3$). Again, we find a uniform distribution $\{1/3,
1/3, 1/3\}$ at high driving levels, and a single-peak distribution at milder driving, but in contrast to the $N = 2$ case the transition between them is hysteretic, see Fig.~\ref{fig-2}. That is, the value of $B$ at which the transition occurs when one goes from vigorous to mild driving is different from the B-value when one goes in the opposite direction. The experimental measurements were 
again done with 600 glass beads ($r$ = 1.25 mm, $e$ = 0.97) in a 3-compartment tube mounted on a shaker.

To account for this experimental finding, we assume that the
approximate expression for the flux (Eq.~\ref{eq-flux}) remains valid also for 3 compartments. In this model, the dynamics of the system is governed by
the following set of equations:
\begin{eqnarray}
\frac{d n_{1}}{d t} = -2F(n_{1})+F(n_{2})+F(n_{3}) , \qquad \textrm{and cyclic permutations.}
\label{eq-dyn}
\end{eqnarray}

The uniform distribution $\{1/3, 1/3, 1/3\}$ always is a solution, but it is stable only for $B<1$. At this value it turns unstable and gives way to a single-peaked solution. This can be seen as follows.

\begin{figure}
\begin{center}
\includegraphics[width=0.75\textwidth]{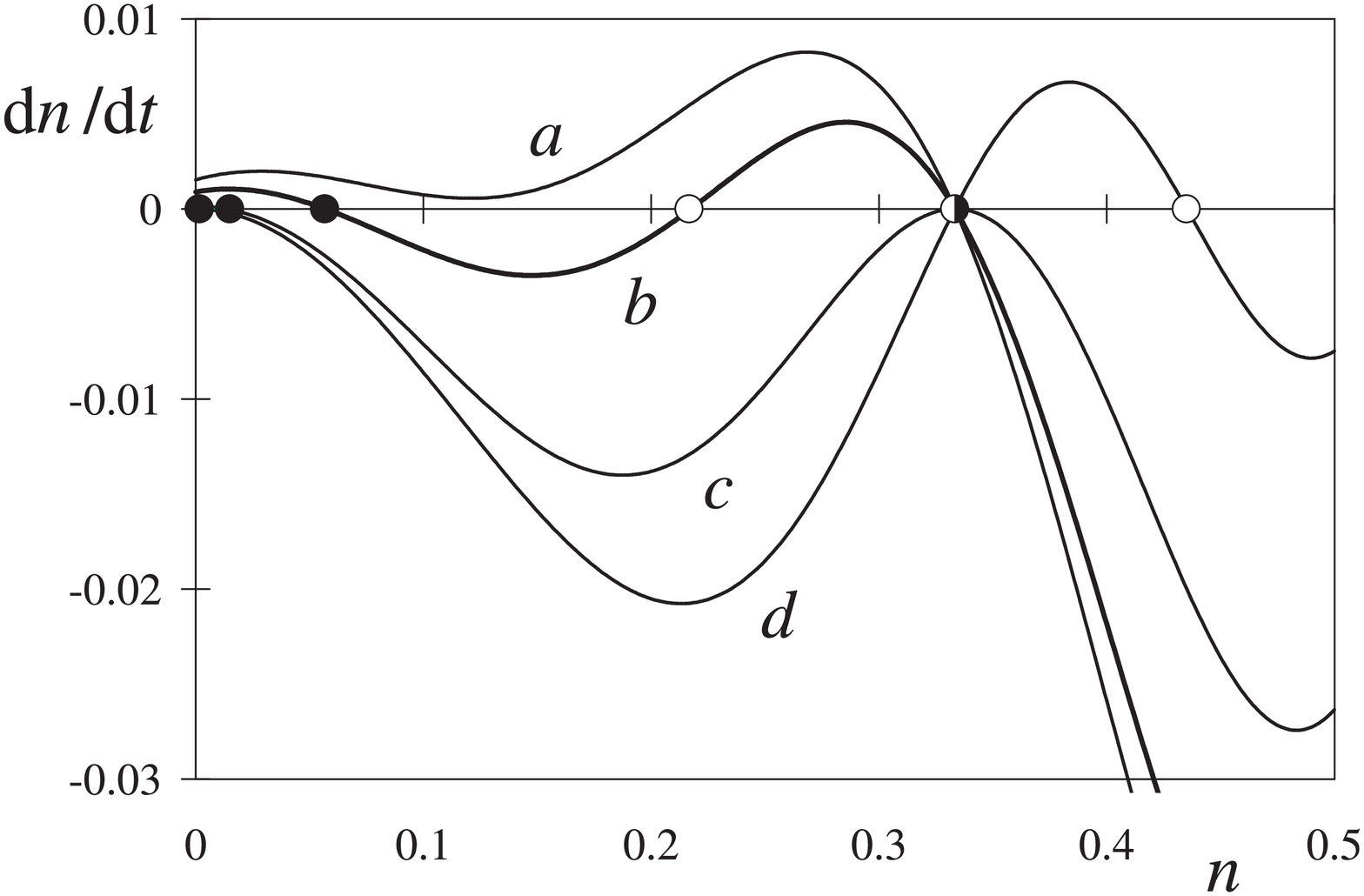}
\includegraphics[width=1.0\textwidth]{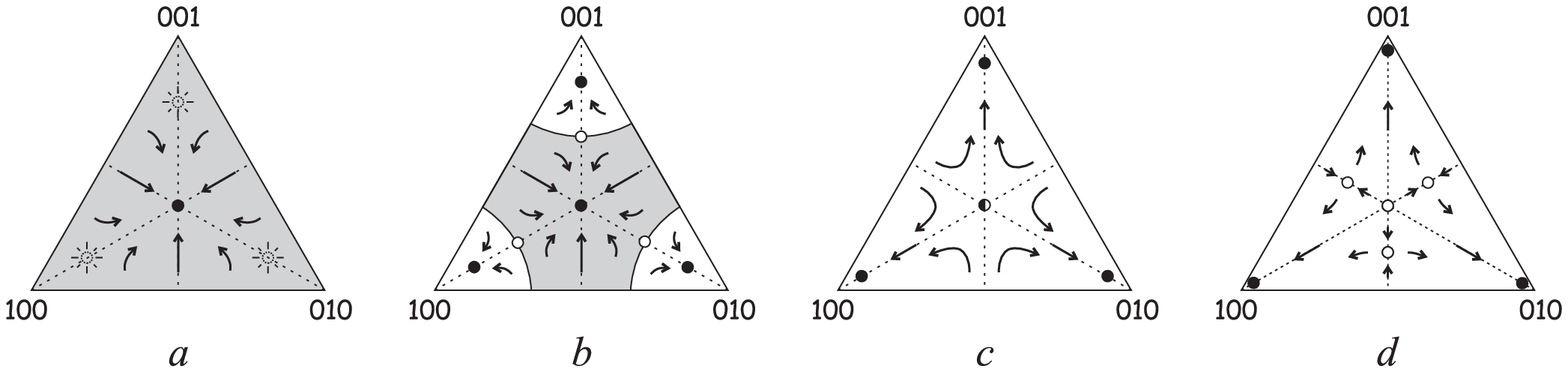}
\caption{The rate of change $dn/dt$ as function of $n$, along the symmetry axes of the 3-box system (cf. Eq. (6)), for four successive values of $B$,
namely,
a) $B=0.72$,
b) $B=0.78$,
c) $B=1.0$,
and d) $B=1.56$.
 For each of these values a triangular flow diagram is given, which shows the dynamics of the full system. The shaded area is the (diminishing) basin of attraction of the uniform solution $\{1/3, 1/3, 1/3\}$, in the center of the triangle;
the white kites are the basins of attraction of the clustering solutions.
Closed circles represent stable equilibria, open circles unstable equilibria.
}
\label{fig-3}
\end{center}
\end{figure}

From a linear stability analysis of the set of Eqs.~\ref{eq-dyn}
around the uniform solution $\{1/3, 1/3, 1/3\}$ we obtain the
eigenvalues 0 and twice $-3F'(1/3)$. The zero eigenvalue, with
eigenvector $(1, 1, 1)$, reflects the fact that synchronously raising
(or lowering) the occupation numbers in the three compartments is forbidden by
particle conservation. The doubly degenerate eigenvalue  becomes
positive for $B>1$, i.e., $\{1/3, 1/3, 1/3\}$ becomes unstable. Any
small perturbations from the symmetric solution will grow, subject of
course to the conservation condition $\sum n_k = 1$, which defines a
plane in the 3D space $\{n_1, n_2, n_3\}$. Indeed, it is exactly this
plane $\sum n_k = 1$ that is spanned by the two eigenvectors belonging
to the degenerate eigenvalue. It is depicted in Fig.~\ref{fig-3} in
the form of a triangle. For instance, the lower left corner
corresponds to the distribution $\{1, 0, 0\}$, and the solid dot in the
center is the uniform distribution $\{1/3, 1/3, 1/3\}$.

Further insight can be gained by direct inspection of Eq.~\ref{eq-dyn}. Since $n_1 = 1 - n_2 - n_3$, this equation can also be written as:
\begin{eqnarray}
\frac{d n_{1}}{d t} = - \frac{d n_{2}}{d t} - \frac{d n_{3}}{d t} =
-2F(1-n_{2}-n_{3})+F(n_{2})+F(n_{3}) , 
\label{eq-dyn2}
\end{eqnarray}
The uniform distribution $( n_1 = n_2 = n_3 )$ is the most symmetric
solution the system admits, and when it bifurcates it gives way to
solutions that necessarily have a lesser degree of symmetry. Numerical
evaluation of Eq.~\ref{eq-dyn} reveals that these new solutions lie on
one of the three (equivalent) lines $n_1 = n_2$, $n_1 = n_3$ and $n_2
= n_3$ (see also Fig.~\ref{fig-3}). We therefore turn our attention to
solutions of Eq.~\ref{eq-dyn2} with the reduced symmetry $n_2 =
n_3$. In the triangular plane $\sum n_k = 1$ of Fig.~\ref{fig-3} this
defines the dashed line from the lower left corner $\{1, 0, 0\}$ to the
middle of the right-hand side of the triangle, $\{0, 1/2, 1/2\}$. With
$n_2 = n_3 \equiv n$, $n_1 = 1 - 2n$, equation~\ref{eq-dyn2} takes the form:
\begin{eqnarray}
\frac{d n}{d t} = F(1-2n) - F(n) = 
A \{ (1-2n)^2 e^{-9B(1-2n)^2} -   n^2 e^{-9Bn^2} \} , 
\label{eq-dyn3}
\end{eqnarray}
and cyclic permutations along the other two symmetry lines. For
increasing values of $B$ the maxima of $F(n)$ and $F(1-2n)$ move in opposite directions, and they go simultaneously through $n=1/3$ at the value $B = 1$. It is precisely at this point that the uniform solution becomes unstable. The situation is depicted in Fig.~\ref{fig-3} for four successive values of $B$.

At $B = 0.72$ (curve \textit{a}) we see that $dn/dt$ has only one zero
(steady state) on the relevant interval $0 \leq n \leq 1/2$, namely,
at $n=1/3$. This solution is stable, as one can easily check from the
sign of $dn/dt$. So regardless of the initial condition one always
ends up in $\{1/3, 1/3, 1/3\}$: its basin of attraction
(the shaded area in Fig.~\ref{fig-3}) is the whole plane $\sum
n_k = 1$. Next, for $B=B_{sn}=0.73$ (not shown), the function
$dn/dt$ touches zero at $n=0.1255$. The index $sn$ denotes that this involves a saddle-node bifurcation.

In curve \textit{b}, at $B = 0.78$, we see that $dn/dt$ has meanwhile
gone through zero, creating one stable and one unstable steady state
along the line $n_2 = n_3$. And, because of the threefold
symmetry of the system, the same has happened along the lines
$n_1 = n_2$ and $n_2 = n_3$. The three newly created stable steady
states are single-peaked distributions. Also the uniform distribution is still stable,
so in the present situation there are four co-existing stable states,
each one surrounded by its own basin of attraction. The three \textit{unstable} states move towards the center of the triangle, closing in upon $\{1/3, 1/3, 1/3\}$ and making its basin of attraction (the shaded area) smaller and smaller for increasing $B$.

Curve \textit{c} depicts the situation for $B = B_{bif} = 1.00$. The three unstable states have just reached the point $\{1/3, 1/3, 1/3\}$, reducing its basin of attraction to zero. At this instant the uniform distribution turns unstable. So from now on all initial configurations end up in one of the three single-peaked distributions.

Curve \textit{d}, at $B = 1.56$, gives an impression of the final
situation. The basins of attraction of the three single-peaked states
divide the triangle into three equal, kite-shaped parts. All the other
steady states are unstable. The three states that have gone through
the point $\{1/3, 1/3, 1/3\}$ are saddle points. Their stable branches
lie along the symmetry lines (cf. the negative slope of $dn/dt$
in Fig.~\ref{fig-3}) and their unstable branches in the perpendicular
directions, as indicated in the triangular plot. So, starting out from the neighborhood of $\{1/3, 1/3, 1/3\}$, one first gets a situation where one box empties itself and two boxes are filled more or less equally, and only later one of these two boxes draws all the beads to itself.

The above sequence of events can be translated into the bifurcation
diagram of Fig.~\ref{fig-2}. The uniform distribution is stable for $0
< B < 1$, and becomes unstable by means of an encounter with the
unstable solutions. These have emerged from the saddle-node
bifurcation at $B_{sn} = 0.73$, together with the stable single-peaked
solutions. The latter solutions remain stable for all $B >
0.73$. In the interval $0.73 < B < 1$ both the uniform and the
single-peaked distributions are stable and, depending on the initial
conditions, the system can end up in either one of them. This
bifurcation structure is very different from the $N=2$ case, where the
transition from uniform to single peaked distribution takes place (see
Fig.~\ref{fig-1}) by means of a simple pitchfork bifurcation, meaning
a 2nd order (continuous) phase transition. The most salient point of
the 1st order (discontinuous) phase transition for $N = 3$ is indeed
the hysteresis, just as found in the experiment. If one gradually
increases the value of $B$ from zero upwards (i.e, if one reduces the
driving), the transition from uniform to single-peaked distribution
takes place at $B_{bif}=1$. If the value of $B$ is then gradually
turned down again, the reverse transition occurs at $B_{sn} = 0.73$.

Physically, this difference stems from the fact that the forward
transition has more degrees of freedom at its disposal than the
reverse one. The former can take place via any path through a 2D
section of the flow diagram (see Fig. 3d) whereas the latter is
confined to take place along one of the 1D symmetry lines (see Fig. 3a). For
the 2-box system (where the flow diagram reduces to a line) there is
no room for any difference of freedom between the forward and backward
transition, and hence there is no hysteresis. 

As for the quantitative agreement between theory and experiment, we
note that the experimental result for the ratio $B_{sn}/B_{bif}$ ($\approx
0.88$ for the results displayed in Fig.~\ref{fig-2}) is larger than the
theoretical value 0.73. This can be explained from statistical
fluctuations in the particle fractions (typically of order $1/\sqrt{P}$, and larger near a bifurcation), which in the neighborhood of $B=1$ extend beyond the rapidly decreasing basin of attraction of the $\{1/3, 1/3, 1/3\}$ state, causing the system to switch prematurely to a single-peaked distribution. This means that in a series of experiments for increasing $B$, the measured bifurcation value $B_{bif}$ will be smaller than the theoretical one. Analogously, in a series for decreasing $B$ (with a single-peaked distribution as initial condition) the measured value for $B_{sn}$ will be somewhat larger than in theory. Therefore, the experimentally determined ratio $B_{sn}/B_{bif}$ should actually be regarded as an upper bound. In addition, of course, it must be recalled that the flux function in Eq.~\ref{eq-flux} is approximate, and that the theoretical value of 0.73 is therefore an approximation,
 too. Any small changes in the function $F(n_k)$ will affect the ratio $B_{sn}/B_{bif}$, however not the qualitative properties of the model.

Not only the hysteresis, but also other features of the model show up in the experiment. In particular, the tendency to move from $\{1/3, 1/3, 1/3\}$ to a single-peaked distribution via a transient saddle-point state with two boxes filled almost equally (leaving the third one empty) is observed. For $B >> 1$ 
(i.e., at very mild driving) the system is even observed to be frozen in this transient state. Both half-filled compartments are so cold that the particles cannot cross the barrier anymore, which in the triangular diagrams corresponds to vanishingly small flux arrows around the saddle point.

In conclusion, we see that the clustering effect in the Maxwell-demon experiment with 3 cyclic compartments is hysteretic. This remains true if the system is made non-cyclic (e.g., by making one of the walls very high, or by putting the boxes in a linear row) and also if one takes more compartments ($N  = 4, 5, 6, ..$). The uniform distribution still becomes unstable at the value $B_{bif} = 1$, but by then there is already a number of alternative solutions. Indeed, for growing $N$ one gets increasingly complex patterns (multi-peak distributions) and the sensitive dependence on initial conditions becomes more prominent.

\acknowledgments
We are grateful to Henni Scholten and Gert-Wim Bruggert for building
the experiment, and to Joris de Vries, Renate Heijmans, and Radboud
Nelissen for measuring the experimental bifurcation diagrams.

\end{document}